\documentclass[12pt]{iopart}
\usepackage{graphicx}
\usepackage[T1]{fontenc}
\usepackage[utf8]{inputenc}
\usepackage{iopams}  
\expandafter\let\csname equation*\endcsname\relax
\expandafter\let\csname endequation*\endcsname\relax
\usepackage{amsmath}
\usepackage{breqn}
\begin{document}

\title[Time-frequency mapping of two-colour photoemission]{Time-frequency mapping of two-colour photoemission driven by harmonic radiation }

\author{Bruno Moio$^{1,2,*}$, Gian Luca Dolso$^1$, Giacomo Inzani$^1$, Nicola Di Palo$^{1,2}$, Rocío Borrego-Varillas$^2$, Mauro Nisoli$^{1,2}$ and Matteo Lucchini$^{1,2}$\footnote[0]{$^*$Author to whom any correspondence should be addressed}}

\address{$^1$ Department of Physics, Politecnico di Milano, 20133 Milano, Italy}
\address{$^2$ Institute for Photonics and Nanotechnologies, IFN-CNR, 20133 Milano, Italy}

\ead{bruno.moio@polimi.it}
\vspace{10pt}
\begin{indented}
\item[]June 2021
\end{indented}

\begin{abstract}
The use of few-femtosecond, extreme ultraviolet (XUV) pulses, produced by high-order harmonic generation, in combination with few-femtosecond infrared (IR) pulses in pump-probe experiments has great potential to disclose ultrafast dynamics in molecules, nanostructures and solids. A crucial prerequisite is a reliable characterization of the temporal properties of the XUV and IR pulses. Several techniques have been developed. The majority of them applies phase reconstruction algorithms to a photoelectron spectrogram obtained by ionizing an atomic target in a pump-probe fashion. If the ionizing radiation is a single harmonic, all the information is encoded in a two-color two-photon signal called sideband (SB). In this work, we present a simplified model to interpret the time-frequency mapping of the SB signal and we show that the temporal dispersion of the pulses directly maps onto the shape of its spectrogram. Finally, we derive an analytical solution, which allows us to propose a novel procedure to estimate the second-order dispersion of the XUV and IR pulses in real time and with no need for iterative algorithms.
\end{abstract}

%

\section{Introduction}\label{intro}

In recent years, ultrashort extreme-ultraviolet (XUV) pulses, produced by high-order harmonic generation (HHG) in gases, in combination with femtosecond infrared (IR) pulses in pump-probe experiments, have unlocked the possibility to study ultrafast dynamics in atoms, molecules and solids with unprecedented temporal resolution, down to the attosecond domain \cite{Krausz2009,Cirelli2018,calegari2014ultrafast,Vos2018,Nisoli2017}. The use of HHG is particularly interesting since it offers the possibility to generate ultrashort pulses, tunable in a broad spectral region from ultraviolet (UV) to soft X-rays. Important physical processes unfolding on a few-femtosecond timescale (\textit{e.g.}, conical intersections in molecules, dynamics of core and deep valence levels, etc.) can be investigated by using XUV pulses, produced by filtering single harmonics (SHs) in a broad harmonic spectrum. This can be obtained, for example, by using a time-delay compensated monochromator \cite{Poletto2007,Poletto2009}. Recently, experiments based on a two-colour pump-probe scheme with SH pulses where used to investigate the dissociate ionization in N$_2$ \cite{eckstein2015dynamics} and the coupled electronic-nuclear dynamics in simple \cite{von2018conical} and complex molecules \cite{reitsma2019delayed,Herve2021}, where the high time resolution, combined with energy tunability, offers new scenarios \cite{Lucchini2019}. In all these experiments a precise assessment of the properties under investigation stems from a proper and reliable characterization of the few-fs pulses. While this task may appear less demanding than the characterization of attosecond pulses, its fulfilment in the XUV regime is not trivial \cite{lucchini2018few}. The temporal properties of both the XUV and IR pulses can be extracted from a SH photoelectron experiment using phase retrieval algorithms \cite{murari2020robustness}. However, assessing the pulses properties directly from the photoelectron spectra, without inversion procedures, would provide a method to intuitively interpret the experimental traces and would be more efficient in terms of computational time.

Here we analyze the time-frequency properties of a two-color, two-photon photoelectron signal, called sideband (SB), obtained by ionizing a target atom with an XUV SH radiation, in the presence of a suitably-delayed IR pulse \cite{glover1996observation}. Unlike previous works, we study the effect of the pulse spectral dispersion beyond the second order, highlighting how the group delay is directly imprinted in the SB time-frequency distribution. By means of a simplified model, we show that the SB signal can be interpreted as a time-frequency distribution of the pulses involved in the photoionization experiment. This confirms that the SBs carry the temporal properties of the pulses in their shape \cite{lucchini2018few}, as for the case of a Gabor-Wigner transform for signal \cite{pei2007relations} and pulse analysis \cite{hong2002time}. Furthermore, in the particular case of second-order dispersion, we show that the SB signal can be analytically calculated, thus enabling a direct estimate of the temporal characteristics of the pulses.

This work is organized as follows: in section~\ref{SHS} we discuss the physical interpretation of a single-harmonic spectrogram (SHS), in section~\ref{FSB} we introduce a simplified model and analyze the properties of the SB signal. Finally, in section~\ref{CON} we draw the conclusions.

\section{Single-harmonic spectrogram (SHS)}\label{SHS}

The photoionization of a rare gas by XUV radiation in the presence of a delayed IR pulse, can be described within the strong field approximation (SFA) by the following expression \cite{kitzler2002quantum} (atomic units are used, unless otherwise specified):
\begin{equation}\label{spectrogramFormula}
S\left(\textbf{p},\tau\right) = \left| \int_{-\infty}^{+\infty} \rmd t X_{si}\left(t+\tau\right) e^{i\phi\left(\textbf{p},t\right)} e^{i\left(\frac{p^2}{2}+I_p\right)t} \right|^2
\end{equation}
where $X_{si}(t)$ is the photoelectron wavepacket which, far from resonances and strong modulations of the species dipole moment \cite{yakovlev2010attosecond, borrego2021reconstruction}, equals the XUV temporal profile, $E_{XUV}(t)$. $I_p$ is the gas ionization potential, while $\textbf{p}$ is the final electron momentum and $\tau$ is the delay between the XUV and IR pulses. In this simplified picture, the XUV pulse produces a photoelectron burst while the IR pulse, described by its vector potential, $\textbf{A}_{IR}$, acts as an ultrafast phase modulator through the term:
\begin{equation}\label{phase}
\phi\left(\textbf{p},t\right) = -\int_{t}^{\infty} \rmd t' \left( \textbf{p} \cdot \textbf{A}_{IR}(t') -  A_{IR}^2(t')/2 \right)
\end{equation}
The collection of photoelectron spectra described by $S\left(\textbf{p},\tau\right)$, is called spectrogram and represents the photoemission probability as a function of $\textbf{p}$ and $\tau$. For big values of $\tau$ the spectrogram coincides with the Fourier transform of the attosecond radiation (rescaled by $I_p$). The evolution of $S$ at small delays depends, instead, on the temporal properties of $E_{XUV}$ and $A_{IR}$. In the case of an isolated attosecond pulse, the photoelectron center of mass follows the IR vector potential, and the spectrogram is called attosecond streaking trace \cite{itatani2002attosecond, mairesse2005frequency}. If the XUV radiation is composed by an attosecond pulse train, the photoelectron spectra at large delays is characterized by discrete peaks spaced by twice the IR photon energy and associated with the direct ionization by the harmonic photons. At small delays, the interaction with the IR field creates additional photoelectron peaks in between the harmonic signal, called sidebands (SBs). In a photon picture, they can be explained in terms of the absorption of an harmonic photon followed by the additional absorption/emission of an IR photon \cite{Paul2001,Muller2002}. Considering a classical IR field as in \eqref{spectrogramFormula}, they can also be understood as the result of the interference among the streaking traces originated by each pulse composing the train \cite{lucchini2015semi,Cattaneo2016}. In this work we will concentrate on the spectrogram originated by a SH, which corresponds to a few-fs XUV pulse in time domain.
\begin{figure}
  \centering
  \includegraphics[width=0.45\textwidth]{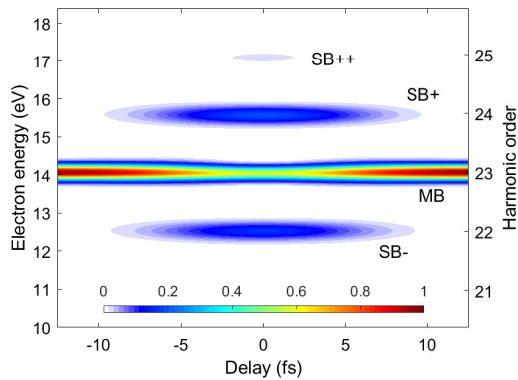}
  \caption{Example of a single-harmonic spectrogram computed with \eqref{spectrogramFormula}. The XUV pulse has a bandwidth of 0.4\,PHz (about 5.9\,fs duration), centered around the 23$^{rd}$ harmonic of 800-nm radiation. The IR has a bandwidth of 0.3\,PHz, corresponding to a pulse of 7.8\,fs, and an intensity of $5 \cdot$ 10$^{11}$\,W/cm$^2$. Both pulses have no dispersion. The noble gas considered in the simulation is neon.}  
  \label{fig:TL}
\end{figure}
Figure~\ref{fig:TL} displays an example of SHS calculated for transform-limited pulses (no spectral dispersion). The IR intensity is $5~\cdot$~10$^{11}$\,W/cm$^2$ and IR and XUV bandwidths are 0.3\,PHz and 0.4\,PHz respectively, corresponding to a full-width-half-maximum in time of about 7.8\,fs and 5.9\,fs. The XUV pulse is centered at 35.65\,eV (i.e. the 23$^{rd}$ harmonic of 800-nm radiation) and the noble gas is neon (I$_p$ = 21.6\,eV). At large delays $S(\textbf{p}, \tau)$ presents a single peak corresponding to direct ionization by the harmonic, hereafter called main band (MB). Near the zero delay (i.e. the temporal overlap of the XUV and IR pulses), SBs appear above and below the MB. Previous works \cite{lucchini2018few,murari2020robustness} proved such SHS to be sensitive to the group delay dispersion of both XUV and IR pulses, demonstrating the applicability of phase reconstruction algorithms like the extended ptychographic iterative engine (ePIE) \cite{Spangenberg2015,lucchini2015ptychographic,murari2020robustness}, to obtain their complete temporal characterization. In the following section we will concentrate on SHS and show how high order dispersion terms are mapped in the delay-energy distribution of the first SB (labelled with SB$^+$ and SB$^-$ in figure~\ref{fig:TL}).

\section{First SB delay-energy distribution}
\label{FSB}

At a given pulse duration, the number of SBs appearing in the SHS depends mainly on the chosen IR intensity. The first SB is formed by a two-color two-photon (XUV and IR) transition. As a result, its delay-energy distribution carries information on both the XUV and IR light temporal characteristics.
\begin{figure*}
  \centering
  \includegraphics[width=1\textwidth]{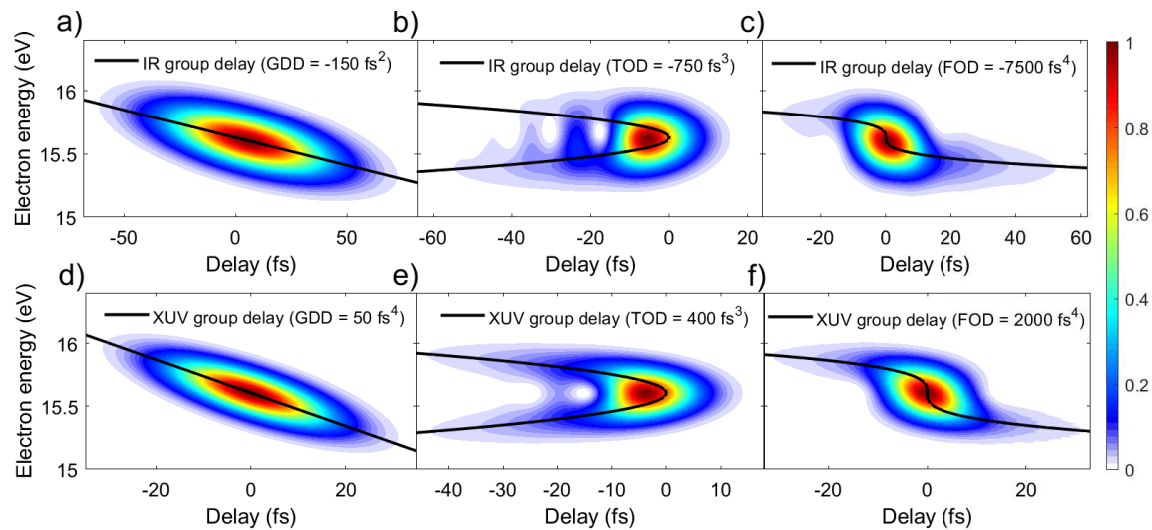}
  \caption{Simulated first upper SB$^+$ for a transform-limited XUV and an IR radiation with second, \textbf{(a)}, third, \textbf{(b)}, and fourth, \textbf{(c)}, order spectral dispersion. \textbf{(d)}-\textbf{(f)} Same quantity but for a transform limited IR and a dispersed XUV. For any order, the shape of the SB$^+$ qualitatively tracks the group delay dispersion of the IR pulse (black solid curve), exhibiting a linear tilt in the case of a GDD, a parabolic distortion in the case of a TOD and a cubic distortion with an FOD. This mimics the behaviour of a time-frequency distribution.}  
  \label{fig:IRdisp}
\end{figure*}
Figures~\ref{fig:IRdisp}(a)-(c) show the effect of a finite IR second, third and fourth order dispersion (GDD, TOD and FOD, respectively) on the first upper SB of the spectrogram of figure~\ref{fig:TL}. Figures~\ref{fig:IRdisp}(d)-(e) show the same analysis, but considering a transform limited IR pulse and a finite dispersion of the XUV pulse. In general, we find the SB to closely track the group delay of either the IR or XUV pulses, marked with a black continuous curve. The effect of a second order dispersion has already been reported by Lucchini \textit{et al.} \cite{lucchini2015ptychographic}: as a result of the linear distribution of the pulse frequencies along their temporal profiles, a finite GDD produces a linearly tilted SB in the delay-energy space. The effect of a finite TOD and FOD is instead reported here for the first time. We found the SB to be profoundly distorted by a finite TOD (figures~\ref{fig:IRdisp}(b) and (e)) or FOD (figures~\ref{fig:IRdisp}(c) and (f)), changing its structure in a characteristic manner. In particular, in the case of a pulse with a TOD, the higher and lower frequencies are shifted on the same side of the pulse envelope, with a quadratic distribution. This produces a c-shaped SB that follows a parabolic energy spreading, as expected from the associated group delay (black curve in the panels). With a FOD, instead, the SB exhibits an s-shaped distribution, tracking a cubic function. The good agreement between the SB delay-energy distribution and the group delay of the pulses can be easily explained with an intuitive picture. As the photoemission involves simultaneous interaction with the two fields, by changing the relative delay between them, we change the portion of pulses which actually overlap in time. Therefore, the time at which different frequencies appear in the pulse, namely the group delay, is mapped onto the final photoelectron distribution. In this regards, one of the pulses can be thought to behave as a gate for the other, as in the case of an ordinary time-frequency distribution \cite{pei2007relations}.

In all the cases discussed above, we never distinguished between the dispersion of the XUV or IR pulse. This is because, apart from a symmetric or antisymmetric SB formation, as reported in figure~\ref{fig:TL} and discussed in the next session, the finite dispersion of the two pulses has the same effect. The main difference resides in the case of third and fourth order dispersion. Here we notice an asymmetry (lower signal at higher energies of the SB) if the IR is dispersed. With a TOD and FOD on the XUV pulse, instead, this effect is not present. However, if we compare the SB shape to the considered group delay, they still match in both cases. Therefore, this effect, which we will address in the next section, does not represent a real limitation to the interpretation we just introduced.

\subsection{Perturbative approach}
\label{STRIPE}

To better understand the intuitive description of the first SB formation we discussed in the previous section, we can develop equations \eqref{spectrogramFormula} and \eqref{phase} to obtain a simplified model to describe the photoionization of a rare gas by a SH in the presence of an IR pulse.

Starting from \eqref{spectrogramFormula}, if we are far enough from atomic resonances and the rare gas dipole moment is sufficiently regular, we can apply the wave packet approximation (WPA) \cite{yakovlev2010attosecond}, which states that the photoelectron wavepacket $X_{si}$ is very well approximated by the XUV electric field $E_{XUV}$ if the ionization cross section of the gas is constant. Given its relatively limited bandwidth, this approximation is justified for a SH. In \eqref{phase}, we can apply the central momentum approximation (CMA), replacing the final electron momentum $\textbf{p}$ by the average momentum of the wavepacket $\textbf{p}_c$. Moreover, we can assume the latter to be oriented along the IR field polarization, such that $\textbf{p}_c \cdot \textbf{A}_{IR} \approx p_c \, A_{IR}$. These considerations are common and widely used when employing the frequency-resolved optical gating for complete reconstruction of attosecond bursts (FROG-CRAB) approach \cite{mairesse2005frequency} to characterize attosecond pulses. At this point, we perform the perturbative approximation, assuming that the IR vector potential has a low intensity ($\lesssim~10^{11}$\,W/cm$^2$). As a result, \eqref{phase} further simplifies and becomes:
\begin{equation}\label{phase2}
\phi\left(t\right) = \int_{t}^{\infty} \rmd t'  p_c \, A_{IR}(t')
\end{equation}
If we express the IR vector potential in terms of its envelope and carrier,
\begin{equation}\label{vecPot}
A_{IR}(t) = A(t) \cos{\left(\omega_0 t + \phi_0\right)},
\end{equation}
we can apply the slowly varying envelope approximation (SVEA) while solving the integral of \eqref{phase2} and write the phase $\phi\left(t\right)$ as:
\begin{equation}\label{phase3}
\phi\left(t\right) \approx - \frac{p_c}{\omega_0^2} \, E_{IR}(t),
\end{equation}
where $E_{IR}(t) = - \frac{\rmd}{\rmd t}\left[A_{IR}(t)\right]$ is the IR electric field. In light of these approximations, we can reformulate \eqref{spectrogramFormula} into the following:
\begin{equation}\label{spectrogramFormulaSTRIPE}
S\left(\omega,\tau\right) \approx \left| \int_{-\infty}^{+\infty} \rmd t E_{XUV}\left(t+\tau\right) e^{i\frac{p_c}{\omega_0^2} \, E_{IR}(t)} e^{i\omega t} \right|^2
\end{equation}
where we set $\omega = p^2/2 + I_p$ to write $S$ in the form of a Fourier transform.

\begin{figure}
  \centering
  \includegraphics[width=0.5\textwidth]{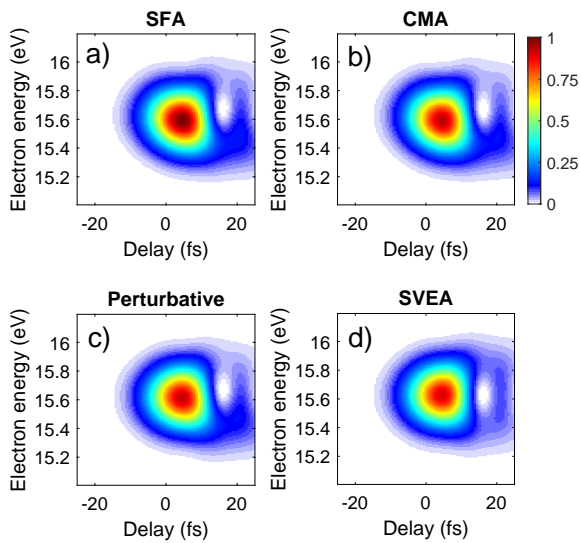}
  \caption{Effect of the consecutive approximations introduced (\textbf{(a)} SFA, \textbf{(b)} CMA, \textbf{(c)} perturbative and \textbf{(d)} SVEA) for the case of a third order dispersion on the IR pulse. The SVEA is responsible for the major distortion of the SB shape, resulting in an up-down symmetrization. This effect is observed only for what concerns the dispersion of the IR pulse and not for a dispersed XUV radiation.}  
  \label{fig:approx}
\end{figure}

In case of relatively long pulses ($\gtrsim$ 3-4 optical cycles), \eqref{spectrogramFormulaSTRIPE} reproduces the photoelectron spectrogram with high degree of fidelity. To test its accuracy, we ran different simulations with an increasing level of approximation, as reported in figure~\ref{fig:approx},
where we display the simulated first-upper SB. In figure~\ref{fig:approx}(a) we show the simulations done with equation~\eqref{spectrogramFormula}, i.e. no approximation applied, whereas figure~\ref{fig:approx}(b) shows the effect of the CMA. Figure~\ref{fig:approx}(c) adds the effect of the perturbative approximation. Finally, figure~\ref{fig:approx}(d) shows the effect of all the approximations, up to the SVEA. The parameters of the simulations are the same as in figure~\ref{fig:TL}, with in addition a third order dispersion of 500\,fs$^3$ on the IR. Such a case corresponds to an intense IR radiation with large dispersion and short TL. We chose the simulation parameters to stress the approximations and test their validity in a bad-case scenario. Despite the approximations of \eqref{spectrogramFormulaSTRIPE} seem very rough, the model still produces a very accurate trace even in this extreme case. In fact, the CMA does not have a large effect on the SB shape (only reduces its amplitude), as well as the perturbative approximation does not affect the structure and the features of the 2D signal. The SVEA is responsible for the main distortion, resulting in an up-down symmetrisation of the SB. However, the distortion has a minor impact on the SB shape and how the temporal information is impressed on it. This proves that \eqref{spectrogramFormulaSTRIPE} is a proper tool for the description of the SB formation in a photoelectron experiment, even in cases where the approximations are at their limits of applicability.

To provide a further prove of this, we can expand the exponential in \eqref{spectrogramFormulaSTRIPE} in its Taylor series, that becomes:
\begin{equation}\label{exponExpanded}
e^{i\frac{p_c}{\omega_0^2} \, E_{IR}(t)} = 1 + i\frac{p_c}{\omega_0^2} E_{IR}(t) + \frac{1}{2} \left( i\frac{p_c}{\omega_0^2} E_{IR}(t)\right)^2 + ...
\end{equation}
If we concentrate solely on two-photon processes, we can neglect the higher-order terms and consider the first one only. The spectrogram is then represented by:
\begin{equation}\label{spectrogramFormulaSTRIPE2}
S\left(\omega,\tau\right) \approx \frac{p_c^2}{\omega_0^4} \left| \int_{-\infty}^{+\infty} \rmd t E_{XUV}\left(t+\tau\right) \, E_{IR}(t) \, e^{i\omega t} \right|^2
\end{equation}
From a physical point of view, this expression describes the first SB formation as sum-frequency event, between an IR photon and an XUV photon \cite{lucchini2018few}. If the photon energy is not evenly distributed in time, the final SB energy changes following the group delay of the two pulses as discussed above. This effect can also be explained from a mathematical point of view because \eqref{spectrogramFormulaSTRIPE2} can be interpreted as a time-frequency distribution, such as the Wigner transform \cite{hong2002time}.

\subsection{Geometrical interpretation of the first SB}

As we described in the previous section, a quadratic phase dispersion produces a tilted SB in the SHS. In particular, a chirped XUV is expected to produce SBs with a parallel delay-frequency tilt, while a non-zero GDD of the IR field makes the SBs to assume a funnel shape (opposite sign of the induced tilt) \cite{lucchini2018few}. The simplified description of \eqref{spectrogramFormulaSTRIPE2} allows for an easy explanation of these effects. 

For simplicity, let's consider two Gaussian pulses of the following complex form:
\begin{eqnarray}\label{newFields}
\fl E_{XUV}(t) = \frac{E_x}{2} \exp{\left[-\frac{t^2}{\gamma_x^2} + i \left(\omega_xt+\eta_x t^2\right)\right]} + cc.
 \nonumber\\ \fl E_{IR}(t) = \frac{E_0}{2} \exp{\left[-\frac{t^2}{\gamma_0^2} + i \left(\omega_0t+\eta_0 t^2\right)\right]} +cc.
\end{eqnarray}
where $E_j$ and $\omega_j$ ($j=x,0$) represent the fields amplitude and frequency. The parameter $\gamma_j$ is related to the transform-limited width of the Gaussian envelope, $\sigma_j$, by the following equation:
\begin{equation} \label{cirpedDuration}
    \gamma_j = \sigma_j\sqrt{1 + \left(\frac{2\beta_j}{\sigma_j^2}\right)^2}
\end{equation}
where $\beta_j$ represents the GDD, defined as the second derivative of the pulse spectral phase $\beta_j~=~\left. \frac{\rmd^2\phi_j(\omega)}{\rmd \omega^2}\right|_{\omega_j}$. With this definition the actual Gaussian envelope full-width-half-maximum duration is given by $FWHM_j~=~\gamma_j\sqrt{2\ln2}$. The parameter $\eta_j$ in \eqref{newFields}, called chirp rate, expresses the quadratic dependence of the instantaneous field frequency over time. For Gaussian pulses it is possible to show that this parameter is linked to $\beta_j$ and $\sigma_j$ by:
\begin{equation} \label{cirpRate}
    \eta_j = \frac{2\beta_j}{\sigma_j^4+4\beta_j^2}
\end{equation}
If we substitute the pulses in \eqref{newFields} into \eqref{spectrogramFormulaSTRIPE2} it is easy to show that the first SB signal originates from the Fourier transform of the term: 
\begin{equation} \label{ChirpedSB}
    \exp{\left\{-\left(\frac{1}{\gamma_x^2} +\frac{1}{\gamma_0^2}\right)t^2 + i \left[\left(\omega_x\pm\omega_0\right)t+\left(\eta_x\pm\eta_0\right) t^2\right]\right\}}.
\end{equation}
While the plus sign corresponds to the absorption of an IR photon associated to the upper SB formation, the minus sign describes the emission of one IR photon which produces the lower SB (respectively SB$^+$ and SB$^-$ in figure~\ref{fig:TL}). From \eqref{ChirpedSB} it is clear that a finite XUV chirp ($\eta_x\neq 0$) has the same effect over the photoelectron central frequency for SB$^+$ and SB$^-$, thus inducing a parallel tilt, while the change of sign is responsible for the opposite funnel-like tilt induced by a non-zero IR chirp ($\eta_0\neq 0$).

For the Gaussian pulses described in \eqref{newFields}, \eqref{spectrogramFormulaSTRIPE2} has an analytical solution which allows to test our model from a quantitative point of view. For the sake of clarity, let's concentrate on the first upper SB only, SB$^+$, and redefine the frequency axis in order to be centered at nominal SB energy, by introducing $\omega' = \omega - \left(\omega_x+\omega_0\right)$. Following equation \eqref{spectrogramFormulaSTRIPE2}, the SB$^+$ signal can be approximated by:
\begin{dmath}\label{firstSBSTRIPE}
      \mathrm{SB}^+\left(\omega',\tau\right)  \simeq S_0\exp{\left\{-2\frac{\left(1-\Theta\right)}{\gamma_0^2}\tau^2 + 
     2\Gamma^2H\Theta\omega_\tau \tau - \frac{\gamma_0^2\Theta}{2}\omega_\tau^2 \right\}} 
\end{dmath}
where we have introduced the quantities:
\begin{eqnarray}
\fl \Gamma^{-2} = \gamma_x^{-2}+\gamma_0^{-2}
\nonumber\\ \fl H = \eta_x+\eta_0
\nonumber\\ \fl \omega_\tau = \omega'-2\eta_0\tau
\\ \fl \Theta = \frac{\Gamma^2}{\gamma_0^2\left(1+\Gamma^4H^2\right)}
\nonumber\\ \fl S_0 = \frac{\pi p_c^2}{\omega_0^{4}} \gamma_0 \Gamma\sqrt{\Theta}\left(\frac{E_x E_0}{4}\right)^2\nonumber.
\end{eqnarray}

The SB shape can be analysed by extracting the contours of the surface represented by equation \eqref{firstSBSTRIPE}. They can be expressed by the following formula, as a function of the pulse parameters in \eqref{newFields},
\begin{equation}\label{eqEll}
A\tau^2 + B\tau\omega' + C\omega^{\prime2} = \mathrm{const.}
\end{equation}
where the canonical form parameters are defined as follows:
\begin{eqnarray}\label{ellipseContours}
\fl A = \gamma_x^2\eta_x^2+\gamma_0^2\eta_0^2 + \frac{1}{\gamma_x^2} + \frac{1}{\gamma_0^2}
\fl \nonumber\\ B = \gamma_x^2\eta_x - \gamma_0^2\eta_0
\fl \\ C = \frac{\gamma_x^2+\gamma_0^2}{4}\nonumber
\end{eqnarray}
This implicit equation, \eqref{eqEll}, defines a bundle of ellipses in the $\left(\omega',\tau\right)$ plane. However, the formal analogy is only partial, since the above mentioned space is not homogeneous, being a frequency-time space. To overcome this apparent limitation, we can redefine the axes as $\tau_n=\tau/\sqrt{\gamma_x^2+\gamma_0^2}$ and $\omega_n=\omega' \sqrt{\gamma_x^2+\gamma_0^2}$, such that the new $\left(\omega_n,\tau_n\right)$ space is dimensionless. The normalization parameter $\sqrt{\gamma_x^2+\gamma_0^2}$ is related to the duration of the energy-integral of the SB, as we will show in section \ref{CONV}. In this new space, the contours in \eqref{eqEll} still represent ellipses, with the following new canonical parameters:
\begin{eqnarray}\label{ellipseParametersNorm}
\fl A_n = A\left(\gamma_x^2+\gamma_0^2\right)
 \nonumber\\ B_n = B
 \nonumber\\ C_n = 1/4
\end{eqnarray}
\begin{figure}
  \centering
  \includegraphics[width=0.5\textwidth]{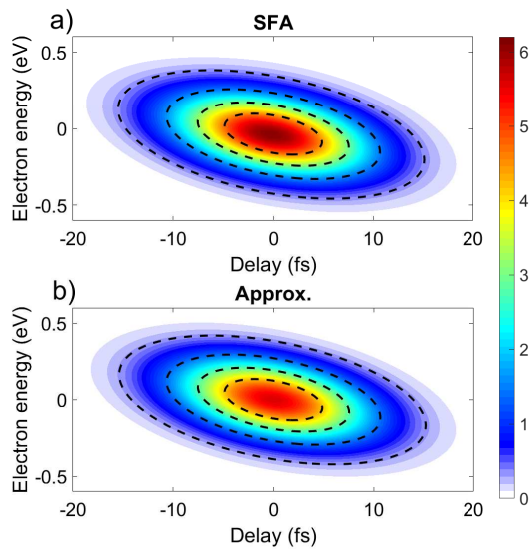}
  \caption{Comparison between SB$^+$ computed with SFA calculations \textbf{(a)} and with the approximated model of \eqref{firstSBSTRIPE} \textbf{(b)}. The approximated model correctly reproduces the SFA results, with only a little discrepancy for what concerns the signal amplitude. In both panels the dashed black ellipses are computed using equation \eqref{eqEll}. They correctly match the shape of the SB.}\label{fig:ellipseComparison}
\end{figure}
In figure~\ref{fig:ellipseComparison} we show a comparison between the first SB, computed with the SFA model (figure~\ref{fig:ellipseComparison}(a)) with the same signal computed with equation \eqref{firstSBSTRIPE}, figure~\ref{fig:ellipseComparison}(b). In both cases, the 2D map is compared to the family of ellipses predicted by equation \eqref{eqEll} (black curve). The XUV and IR  transform-limit durations are 10\,fs and 8\,fs, while the chirp rates are $\eta_x = 0.02$\, fs$^{-2}$ and $\eta_0 = 0.01$\,fs$^{-2}$. The IR intensity is set to 10$^{11}$\,W/cm$^2$. The results highlight the very good agreement between the SFA calculations and the approximation of equation \eqref{firstSBSTRIPE}. Apart from a small error in the amplitude, the geometrical features of SB$^+$ are correctly retrieved.

Once the analogy between the two descriptions is proved, we can discuss in detail the effect of the pulse dispersion on the SB tilt. One way to do so, is to compute the local maxima of the lineouts, extracted at each delay or at each energy. These values can be calculated from equation \eqref{firstSBSTRIPE}, by computing the partial derivatives of the SB with respect to the frequency $\omega'$, delay per delay, or with respect to the delay $\tau$, energy by energy. The result is a linear function in both cases, whose slope is represented by the parameters $m_\omega^+$ and $m_\tau^+$, as described in the following equations:
\begin{eqnarray}\label{slopesPlus}
\fl m_\omega^+ = -2\frac{\gamma_x^2\eta_x-\gamma_0^2\eta_0}{\gamma_x^2+\gamma_0^2}
 \nonumber\\ m_\tau^+ = -2\frac{\gamma_x^2\eta_x^2 + \gamma_0^2\eta_0^2 + \gamma_x^{-2} + \gamma_0^{-2}}{\gamma_x^2\eta_x-\gamma_0^2\eta_0}
\end{eqnarray}
Here, the subscripts $\omega$ or $\tau$ denote the directions along which the maxima are calculated, whereas the superscript "+" indicates the first upper SB. If we consider the lower SB, the IR chirp rate changes its sign, meaning that the slopes are now represented by:
\begin{eqnarray}\label{slopesMinus}
\fl m_\omega^- =  m_\omega^+ \cdot \frac{\gamma_x^2\eta_x+\gamma_0^2\eta_0}{\gamma_x^2\eta_x-\gamma_0^2\eta_0}
 \nonumber\\ m_\tau^- = m_\tau^+ \cdot \frac{\gamma_x^2\eta_x-\gamma_0^2\eta_0}{\gamma_x^2\eta_x+\gamma_0^2\eta_0}
\end{eqnarray}
\begin{figure}
  \centering
  \includegraphics[width=0.5\textwidth]{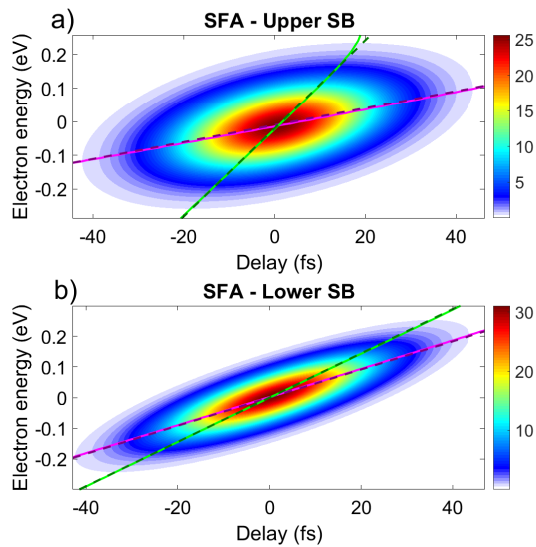}
  \caption{SFA simulation of the upper, \textbf{(a)}, and lower, \textbf{(b)}, SBs corresponding to an XUV transform-limited duration duration of 10\,fs with $\beta_x = 120$\,fs$^2$, an IR transform-limited duration of 12\,fs with $\beta_0 = 50$\,fs$^2$ and an intensity of 10$^{11}$\,W/cm$^2$. The purple and green solid lines represent the numerical maxima of the SB lineouts, per delay and per energy, respectively. The black dashed lines represent the same maxima, computed with slope formulas in \eqref{slopesPlus}.}\label{fig:slopes}
\end{figure}
Figure~\ref{fig:slopes} shows SB$^+$ and SB$^-$ computed with the SFA model, compared to the functions of the maxima as extracted from the SB themselves (purple and green solid lines), and calculated with \eqref{slopesPlus} and \eqref{slopesMinus}. Within the numerical accuracy of the calculations, the analytical prediction properly follows the numerical maxima. 

Besides explaining the universal behaviour of the SB slope numerically observed in Ref.~\cite{lucchini2018few}, the expressions we derived suggest a way to directly estimate the pulse GDDs from the experimental SB tilt, with very little prior information and with no need for iterative procedures. In fact, if we consider the slopes $m_\tau^+$ and $m_\tau^-$ and combine them to \eqref{cirpRate} and \eqref{cirpedDuration}, we obtain the following relation for the pulses GDD:
\begin{eqnarray}\label{slopesDispersion}
\fl \beta_x = \frac{1}{2}\left(1+\frac{\sigma_x^2}{\sigma_0^2}\right)\left(\frac{1}{m_\tau^-}+\frac{1}{m_\tau^+}\right)
 \nonumber\\ \beta_0 = \frac{1}{2}\left(1+\frac{\sigma_0^2}{\sigma_x^2}\right)\left(\frac{1}{m_\tau^-}-\frac{1}{m_\tau^+}\right)
\end{eqnarray}
Therefore, by knowing the transform limited time duration of the pulses (equivalently, their bandwidths) and measuring the SB tilt, it is possible to quickly retrieve the spectral chirp for both IR and XUV. To test this approach, we applied the procedure to two SFA calculations reported in figure~\ref{fig:slopes} and retrieved $\beta_x' = 120.8 \pm 0.3$\,fs$^2$ and $\beta'_0 = 50.2\pm 0.4$\,fs$^2$ which nicely compare with the exact values of $\beta_x = 120.0$\,fs$^2$ and $\beta_0 = 50.0$\,fs$^2$ used in the simulations. As the equations \eqref{slopesDispersion} are supposed to fail for relatively high IR intensities, we repeated the test starting from an SFA trace computed for 10 times higher IR intensity (10$^{12}$\,W/cm$^2$), thus close to the limit of applicability of the simplified model. In this second case we found $\beta_x' = 113 \pm 12$\,fs$^2$ and $\beta_0' = 69\pm20$\,fs$^2$. While the accuracy is reduced, the GDDs are almost correctly recovered, showing that the proposed method is robust with respect to the crude approximation introduced. This proves that the shape of the SBs can be reliably used to characterize the pulses involved in the photoelectron experiment, with no need for complex reconstruction procedures.

\subsection{Convolution properties of the first SB}\label{CONV}

Using equation \eqref{firstSBSTRIPE} we can derive other important relationships between the temporal properties of the pulses and the resulting SB time-frequency distribution. From \eqref{spectrogramFormulaSTRIPE2} it follows that the SB can be seen as the Fourier transform of the product between a signal and a delayed gate, namely:
\begin{equation}\label{spectrogramFormulaSTRIPEfourier}
S\left(\omega',\tau\right) \approx \frac{p_c^2}{\omega_0^4} \left| \mathcal{F} \left\{ E_{XUV}\left(t\right) \, E_{IR}(t-\tau) \right\} \right|^2.
\end{equation}
If we now  we consider the complex representation of the fields reported in equation~\eqref{newFields} (\textit{i.e.} obtained by neglecting the complex conjugate in their definition) by using the convolution property of the Fourier transform in \eqref{spectrogramFormulaSTRIPEfourier}, it is possible to demonstrate that for the photoelectron spectrum at zero time delay, $S(\omega',0)$, it holds:
\begin{equation}\label{spectrogramFormulaSTRIPEfourier1}
S\left(\omega',0\right) \approx \frac{p_c^2}{\omega_0^4} \left| \hat{E}_{XUV}\left(\omega'\right) * \hat{E}_{IR}\left(\omega'\right) \right|^2
\end{equation}
where the symbol $*$ denotes the convolution, while $\hat{E}_{XUV}\left(\omega\right)$ and $\hat{E}_{IR}\left(\omega\right)$ are the Fourier transforms of the complex fields, $E_{XUV}'$ and $E_{IR}'$\\
In a similar way, using the integration property of the Fourier transform in \eqref{spectrogramFormulaSTRIPEfourier}, we can relate the SB lineout at $\omega' = 0$, i.e. at $\omega = \omega_x \pm \omega_0$, to the time convolution of the two complex light pulses: 
\begin{equation}\label{spectrogramFormulaSTRIPEfourier2}
S\left(0,\tau\right) \approx \frac{p_c^2}{\omega_0^4} \left| E'_{XUV}\left(t\right) * E'_{IR}\left(t\right) \right|^2.
\end{equation}
It is worth mentioning that \eqref{spectrogramFormulaSTRIPEfourier1} and \eqref{spectrogramFormulaSTRIPEfourier2} are derived from the mathematical properties of the Fourier transform and thus they are valid in general, regardless of the particular shape of the pulses.

Besides some particular lineouts, it might be interesting to evaluate the SB integrals over delay and frequency. The energy integral of the first upper SB in \eqref{firstSBSTRIPE} yields:
\begin{equation}\label{eneIntegral}
S_\omega = \int_{-\infty}^{+\infty}S\left(\omega,\tau\right) \rmd \omega = A_s\exp{\left(-\frac{2\tau^2}{\gamma_x^2+\gamma_0^2}\right)}
\end{equation}
where $A_s$ is the amplitude, expressed as:
\begin{equation}
A_s = \frac{\sqrt{2\pi^3} p_c^2}{\omega_0^{4}}\left(\frac{E_0E_x}{4}\right)^2 \left(\frac{1}{\gamma_x^2}+\frac{1}{\gamma_0^2}\right)^{-\frac{1}{2}}.
\end{equation}
The integral is therefore a Gaussian function, whose amplitude and standard deviation depend on the actual durations of the pulses. In particular, the width of $S_\omega$ is equal to the width of the convolution of the real field intensity profiles $I_{IR}(t) = \left| E_{IR}\left(t\right) \right|^2$ and $I_{XUV}(t) = \left| E_{XUV}\left(t\right) \right|^2$, explaining why the energy integral of the SB signal scales with the convolution law \cite{lucchini2018few}.

The integral of the SB$^+$ with respect to the delay axis gives instead:
\begin{equation}\label{tauIntegral}
S_\tau = \int_{-\infty}^{+\infty}S\left(\omega,\tau\right) \rmd \tau \propto \exp{\left(-\frac{\omega^2}{2\left(\frac{1}{\sigma_x^2}+\frac{1}{\sigma_0^2}\right)}\right)}.
\end{equation}
If we define the spectral intensities $\hat{I}_{IR}(t) = \left| \hat{E}_{IR}\left(t\right) \right|^2$ and $\hat{I}_{XUV}(t) = \left| \hat{E}_{XUV}\left(t\right) \right|^2$, $S_\tau$ has the same width as their convolution. 
\begin{figure}
  \centering
  \includegraphics[width=0.5\textwidth]{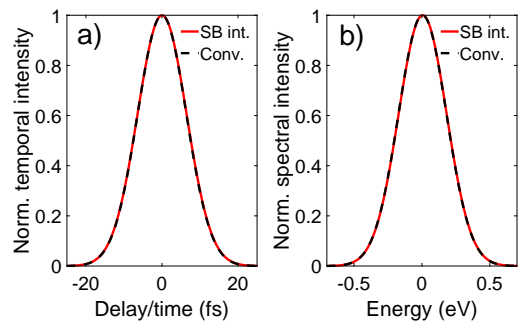}
  \caption{Comparison between the SB integral (red solid curve) and the pulse intensity convolutions (black dashed curve), in the temporal, \textbf{(a)}, and spectral domain, \textbf{(b)}. The remarkable agreement between the curves shows that the SB integrals encode the pulse convolutions.}\label{fig:conv}
\end{figure}
Figure~\ref{fig:conv} compares the SB integrals performed numerically over the SFA calculations of figure~\ref{fig:ellipseComparison}(a) (red solid curves), and estimated with convolution of the pulses temporal and spectral intensities (black dashed curves). The results show good agreement both for the energy integral (figure~\ref{fig:conv}(a)) and the delay integral (figure~\ref{fig:conv}(b)), proving that the SB integrals encode the convolutions of the pulses, both in the temporal and spectral domain.

\section{Conclusions}\label{CON}

We investigated in detail the effect of high-order dispersion terms upon the SB formation in a two-color photoemission process. By means of SFA simulations we showed that the energy-delay distribution of the SB signal qualitatively follows the GDD of the pulses as it generally happens in time-frequency distributions. Indeed, we showed that within the SVEA and the perturbative approximation, the first SB signal can be written as the Fourier transform of the time product of a signal and a gate as in a generalized Gabor transform. In the particular case of Gaussian pulses with a quadratic dispersion, we derived an analytical description of the SB signal whose contour lines are represented by ellipses. This allowed us to put the geometrical properties of the SB signal in direct relation with the temporal properties of the pulses and explain, for example, the different effect of the IR and XUV chirp on the SB tilt. In this view, we showed that one can use the information about the tilt of the upper and lower SBs to simultaneously extract the dispersion of the XUV and IR pulses. Moreover, we showed that the SB profiles at zero pump-probe delay or at the SB central energy, correspond, respectively, to the temporal and frequency convolution of the IR and XUV complex pulses, whereas the energy and delay integrals correspond to the convolution of the intensity temporal and spectral profiles. By fully deriving the chirp dependence of the SB tilt, our results explain why SHSs can be used in a FROG-CRAB approach to retrieve the temporal characteristics of the IR and XUV pulses. Furthermore, they define the applicability of the cross-correlation formula to estimate the SB duration or, equivalently, to get a first estimate of the XUV and IR properties from an experimental SHS. Finally, by giving a more direct picture of the two-color/two-photon photoemission spectra, our findings deepen the comprehension of attosecond photoelectron spectroscopy techniques which allows for an interpretation of the experimental results under a different, and new perspective.

\section*{Funding}
This project has received funding from the European Research Council (ERC) under the European Union’s Horizon 2020 research and innovation programme (grant agreement No.~848411 title AuDACE). ML and GI further acknowledge funding from MIUR PRIN aSTAR, Grant No.~2017RKWTMY. MN acknowledges funding from MIUR PRIN, Grant No.~20173B72NB.

\section*{Data availability statement}
The data that support the findings of this study are available upon reasonable request from the authors.

\section*{References}
\bibliography{references_custom}
\end{document}